\documentclass[3p,twocolumn]{elsarticle}

\usepackage{amssymb}
\usepackage{amsmath}
\usepackage{lineno}
\journal{Chaos Solitons and Fractals}

\begin{document}

%\linenumbers

\begin{frontmatter}

\title{Gradual learning supports cooperation in spatial prisoner's dilemma game}

\author[label1]{Attila Szolnoki}
\author[label2]{Xiaojie Chen}

\address[label1]{Institute of Technical Physics and Materials Science, Centre for Energy Research, Hungarian Academy of Sciences, P.O. Box 49, H-1525 Budapest, Hungary}

\address[label2]{School of Mathematical Sciences, University of Electronic Science and Technology of China, Chengdu 611731, China}

\begin{abstract}
According to the standard imitation protocol, a less successful player adopts the strategy of the more successful one faithfully for future success. This is the cornerstone of evolutionary game theory that explores the vitality of competing strategies in different social dilemma situations. In our present work we explore the possible consequences of two slightly modified imitation protocols that are exaggerated and gradual learning rules. In the former case a learner does not only adopt, but also enlarges the strategy change for the hope of a higher income. Similarly, in the latter case a cautious learner does not adopt the alternative behavior precisely, but takes only a smaller step towards the other's strategy during the updating process. Evidently, both scenarios assume that the players' propensity to cooperate may vary gradually between zero (always defect) and one (always cooperate) where these extreme states represent the traditional two-strategy social dilemma. We have observed that while the usage of exaggerated learning has no particular consequence on the final state, but gradual learning can support cooperation significantly. The latter protocol mitigates the invasion speeds of both main strategies, but the decline of successful defector invasion is more significant, hence the biased impact of the modified microscopic rule on invasion processes explains our observations.
\end{abstract}

\begin{keyword}
cooperation \sep imitation \sep prisoner's dilemma
\end{keyword}

\end{frontmatter}

\section{Introduction}
\label{intro}

There are several update rules which are used in evolutionary game theory where different strategies compete within the framework of a social dilemma \cite{szabo_pr07, perc_bs10, fotouhi_rsif19, takeshue_epl19, cheng_f_pa19, liu_dn_pa19}. In case of birth-death protocol, for instance, a player is chosen from the entire population proportional to fitness and its offspring replaces a randomly chosen neighbor \cite{ohtsuki_jtb06}. Alternatively, in death-birth updating an actor is chosen to die randomly and surrounding neighbors compete for the available site proportional to their fitness. In the best-response dynamics the frequency of a strategy increases only if it is the best response to the actual state \cite{matsui_jet92}. Further possibilities are the so-called win-stay-lose-shift move \cite{nowak_n93, chen_xj_pa08, amaral_pre16, fu_mj_pa19} and the aspiration-driven update when a player's willingness to change depends on a predefined threshold value \cite{liu_yk_epl11, wang_z_pre10, chen_ys_pa17, wu_t_njp18, liu_rr_amc19, zhang_lm_epl19}.

Beside these, the most frequent microscopic dynamics is the celebrated imitation rule where a player may adopt the strategy of a neighbor depending on their payoff difference. The popularity of the latter protocol is based on the fact that not only biological systems but also human behavior can be modeled realistically by this rule \cite{fu_prsb11, naber_pnas13, zhang_by_expecon14}. One of the inspiring observations of previous studies was that the evolutionary outcome may depend sensitively on the applied microscopic dynamics \cite{roca_plr09,du_wb_ijmpc10, szolnoki_njp18b, hadjichrysanthou_dga11, chica_cnsns19, takesue_pa19, szolnoki_pa18}.

Motivated by this phenomenon, in the present work we study how the final solutions change if we slightly modify the standard imitation rule. This question was also inspired by the fact that accurate strategy imitation of a partner is not necessarily a realistic approach in several cases. For example, one can easily imagine a situation when a player, who witnesses the success of a competitor player, does not simply copy the attitude of the other, but tries to surpass the other's behavior for an even higher income. But we can also give examples for an alternative player's approach. In the latter case a careful player is reluctant to execute drastic change from the original strategy. Instead, this actor makes just a minimal step toward the targeted strategy via the update process. 

The above mentioned modified imitation scenarios are studied in the framework of a spatial prisoner's dilemma game where cooperator and defector strategies compete. According to the dilemma the latter strategy provides higher individual income for a player independently from the other's choice while by choosing the former strategy simultaneously competitors would gain a higher mutual income.

Evidently, to adjust the strategy change in different ways involves a multi-strategy system where players can choose not only from two options. This can be easily done if we allow actors to choose intermediate levels of cooperation by assuming continuum of strategies between the two extremes \cite{frean_jtb96}. For easier numerical treatment we propose a spatial prisoner's dilemma game with discrete strategies in order to reveal the possible consequence of inaccurate imitation protocols. Hence, instead of the infinite number of continuous strategies, we introduce $N$ intermediate discrete strategies between the always defector and always cooperator strategies. In this setup the above described protocols are feasible because players' strategy can change gradually.

In the next section we provide the details of our models that is followed by the result section where we present all observed consequences of the modified imitation rules. In the last section we discuss their potential implications and some future research directions are also given.

\section{Gradual propensity level for cooperation}
\label{def}

Starting from the traditional spatial prisoner's dilemma we assume that players are arranged on a spatial graph where every player interacts with their neighbors and gains payoff according to the prisoner's dilemma parametrization. Initially we present results obtained on $L \times L$ square grid with periodic boundary conditions, but the key observations are also verified on other graphs including highly irregular scale-free network and random graph that exhibits small-world character. 

Without loss of generality we assume that the mutual cooperation pays $R=1$ reward for both players, while a defector player gains $1<T<2$ temptation value against a cooperator where the latter sucker's payoff is $S=0$. Finally, mutual punishment leads to $P=0$ individual income. As it was earlier demonstrated this so-called weak prisoner's dilemma parametrization can capture faithfully the key character of a social dilemma \cite{nowak_n92b}.

In the traditional setup the imitation dynamics during an elementary Monte Carlo step is the following. First, a randomly selected player
$x$ having strategy $s_x$ acquires its payoff $P_x$ by playing the game with all its neighbors determined by the interaction graph. After, a neighbor $y$ is chosen randomly whose $P_y$ calculated similarly. Last, player $x$ adopts the strategy $s_y$ from player $y$ with a probability determined by the Fermi function
$W = 1/\{1+\exp[(P_x-P_y)/K]\}$, where $K$ determines the uncertainty of proper imitation \cite{szabo_pr07}. Evidently, if $K$ is high, the imitation happens randomly while in the $K \to 0$ limit the update is deterministic and happens only if $P_y$ exceeds $P_x$. According to the standard simulation technique each full Monte Carlo step gives a chance for every player to change its strategy once on average.

\begin{figure}[h!]
\centering
\includegraphics[width=6.5cm]{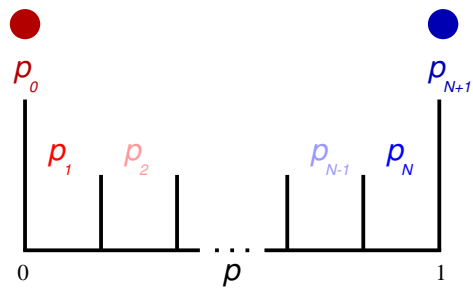}\\
\caption{Coarse grained profile to characterize players' propensity to cooperate. Here we use $N+2$ discrete strategies where the intermediate $N$ strategies covers the whole $(0,1)$ interval uniformly. In particular, a player represented by $p_0$ always defects, while a player belonging to $p_{N+1}$ class always cooperates. In case of an intermediate strategy a player from $p_i$ class cooperates with a probability $(i-1)/N < p < i/N$.}\label{gradual}
\end{figure}

In our present work there are two significant differences from the traditional model. First, the proposed learning protocols can only be executed if we assume that players' propositions to cooperate with others go beyond the traditional two-state setup. In particular, we consider that beside the unconditional cooperate and unconditional defect states a player may cooperate with a certain $p$ probability. Instead of the infinite number of continuous strategies we assume that players can be classified into $N+2$ discrete classes. As Fig.~\ref{gradual} illustrates, beside the traditional $p=0$ (always defect) and $p=1$ (always cooperate) cases we introduce $N$ additional strategies distributed uniformly in the $(0,1)$ interval. For example when a player adopts a strategy belonging to $i$-th intermediate class then it cooperates with others with a probability $p_i$, where $(i-1)/N < p_i < i/N$. Importantly, if $p_i$ and $p_j$ are from the same $[(i-1)/N,i/N]$ interval then player $i$ and $j$ are considered to belong to the same strategy class.

In the following we study two conceptually different imitation protocols. In the first case when a strategy update is executed, a player $i$, whose strategy belongs to the $p_i$ class, does not imitate the $p_j$ strategy of player $j$ directly. Instead player $i$'s new strategy belongs to the class of $p_{j+1}$ if $j>i$ or the new strategy belongs to the class of $p_{j-1}$ when $j<i$. This protocol describes a situation when a learner tries to go beyond the "master" by exceeding latter's strategy choice. In this way a learner hopes an even higher income from the strategy update. Evidently, if $j=N+1$ for $j>i$ then player $i$ accepts $p_{N+1}$ strategy. Similarly, when $j<i$ and $j=0$ then player $i$ adopts $p_0$ strategy. As an important technical detail, when a player adopts an intermediate strategy class $j$ then its new $p_j$ value is generated randomly from the $[(j-1)/N, j/N]$ interval. In this way we can avoid that the initial distribution of $p_i$ values influences the final evolutionary outcome artificially.

According to the alternative imitation protocol the learner player is more careful. Practically it means that when player $i$ changes strategy in response to the success of player $j$ then player $i$'s new strategy is $p_{i+1}$ if $j>i$ or $p_{i-1}$ if $j<i$. In other words, player $i$ is reluctant to change its attitude and follows player $j$ only in a minimal way. Similarly to the previous protocol when an update is executed then a new $p_i$ value is generated from the new interval. We must note that in both cases in the $N=0$ limit we get back the traditional two-strategy model, while for large $N$ values the strategy change can be modified smoothly. 

For the shake of appropriate comparison with the results of traditional model obtained previously we use $K=0.1$ noise level, but our key observations remain intact for other noise values \cite{szabo_pre05, szolnoki_epjb08}. When square lattices were used the linear size was ranged from $L=100$ to $L=6400$. In case of random graph the same $k = 4$ degree distribution was introduced for proper comparison \cite{szabo_jpa04}, where the typical system contained $6 \cdot 10^5$ players. Similarly, for the scale-free graph we kept the average $k = 4$ degree where typical system size was comparable to the random graph. Importantly, in the scale-free network case we applied degree-normalized payoff values to avoid the additional effect of highly heterogeneous interaction graph \cite{santos_prl05}. As it was shown previously for the traditional two-strategy model, when payoff values are normalized by the degree of each player then the average cooperation level decreases drastically and becomes comparable to those values obtained for regular graphs and cooperators cannot survive for $T \ge 1.1$ temptation values \cite{szolnoki_pa08,masuda_prsb07,tomassini_ijmpc07}.

\section{Results}

\begin{figure}
\centering
\includegraphics[width=7.0cm]{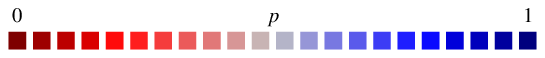}\\
\includegraphics[width=7.8cm]{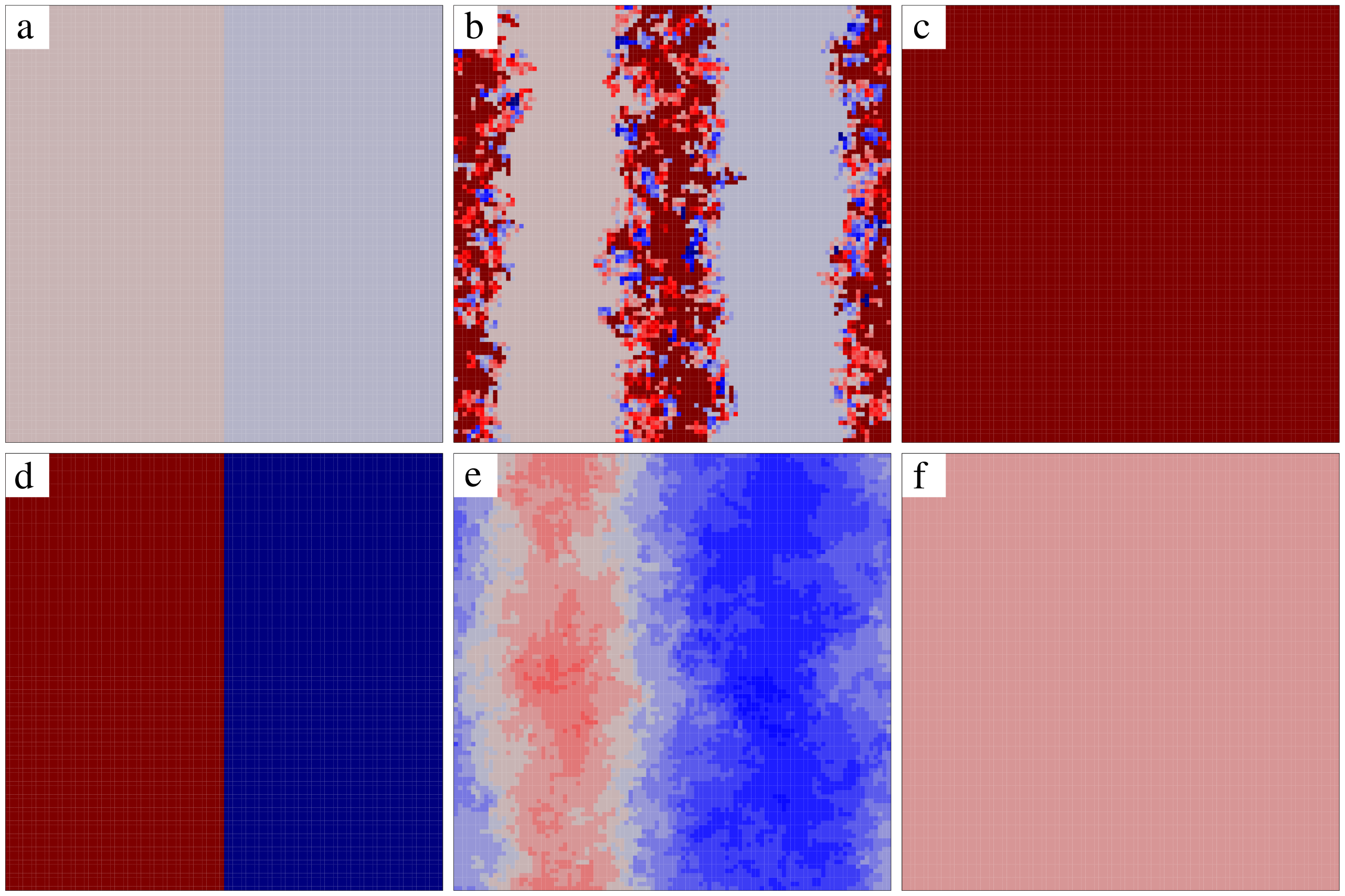}\\
\caption{Front propagation for $N=20, T=1.2$ by using the suggested imitation protocols. Strategies are represented by different colors from deep red (full defection) to deep blue (full cooperation) as indicated. Top panels from (a) to (c) show how the system evolves from two intermediate strategies in case of exaggerated learning protocol. When we launch the evolution from the mentioned prepared initial state the extreme strategies appear very soon and the system evolves to the traditional two-strategy model. Here, due to high temptation value, full defectors prevail. Bottom panels from (d) to (f) show the evolution when gradual learning is applied. Here the initial state contains only the extreme strategies, but the system gradually evolves into a homogeneous state where players willingness to cooperate is significant. The system size is $L=100$ for both cases.}\label{over_under}
\end{figure}

Before presenting the results of modified imitation protocols, it is important to stress that the introduction of multi-strategies to the original model, where not just the extreme but also intermediate strategies compete, do not modify the evolutionary outcome. If a learner player follows the teacher's strategy accurately during the imitation process then the system will gradually terminate onto the classic two-strategy state where multi-strategies become redundant. This phenomenon is nicely illustrated in the animation provided in \cite{accurate}. 
Here the initial state contains 22 strategies ($N=20$) and the temptation value is $T=1.02$. Different strategies are denoted by different shade of red and blue as indicated on the top of Fig.~\ref{over_under}. Accordingly, the deep red marks always defect ($p_0$) strategy, while deep blue marks always cooperate ($p_{21}$) strategy.
When the evolution is launched a coarsening process starts and intermediate strategies die out eventually. Finally only the two extreme strategies survive who form a stable coexistence and provide an $f_C \approx 0.4$ cooperation level in agreement with the result of the traditional model \cite{szabo_pre05}. Evidently, if temptation value is increased then the fraction of cooperators decays and they cannot survive above $T=1.038$.

In the following we compare the consequences of modified imitation protocols by using the same parameter values ($N=20$ and $T=1.2$). Figure~\ref{over_under} highlights that the evolutionary outcomes are strikingly different. When the strategy change is enlarged we started the evolution from a state when only the middle strategies, $p_{10}$ and $p_{11}$ are present. This case is illustrated in the top row from panel~(a) to (c). As panel~(b) shows, the extreme strategies, such as $p_0$ and $p_{21}$, emerge at very early stage of the evolution. Due to the character of the imitation rule intermediate strategies drift toward the edges of profile interval leaving only extreme strategies alive. At such a high temptation level, however, the two-strategy model terminates into a full defector state as shown in panel~(c). The related animation starting from the prepared initial state to the final absorbing state can be seen in \cite{exaggerated}. Based on this the final conclusion can be summarized as the following. Because of the specific feature of exaggerated imitation rule intermediate strategies cannot coexist long and players gradually adopt the extreme strategies. From this point the final outcome is basically determined by the relation of the always defect and the always cooperator strategies. In other words, the modification form accurate imitation to exaggerated strategy change does not alter notably the results observed for the traditional two-strategy model.

The impact of gradual imitation rule, however, is significantly different. This case is illustrated in bottom row of Fig.~\ref{over_under}. Initially, plotted in panel~(d), only extreme strategies are present. When the evolution is launched, intermediate strategies emerge and they gradually invade the homogeneous extreme domains. We observed that full defector island is more sensitive and shrinks faster than cooperator domain. This stage is illustrated in panel~(e). Interestingly, intermediate strategies invade each others, too, and finally only a single strategy survives, which provides an average $f_C=0.4$ cooperation level for the whole population. The final state is shown in panel~(f) and the related animation of the whole process can be seen in \cite{gradual}.

\begin{figure}[h!]
\centering
\includegraphics[width=7.0cm]{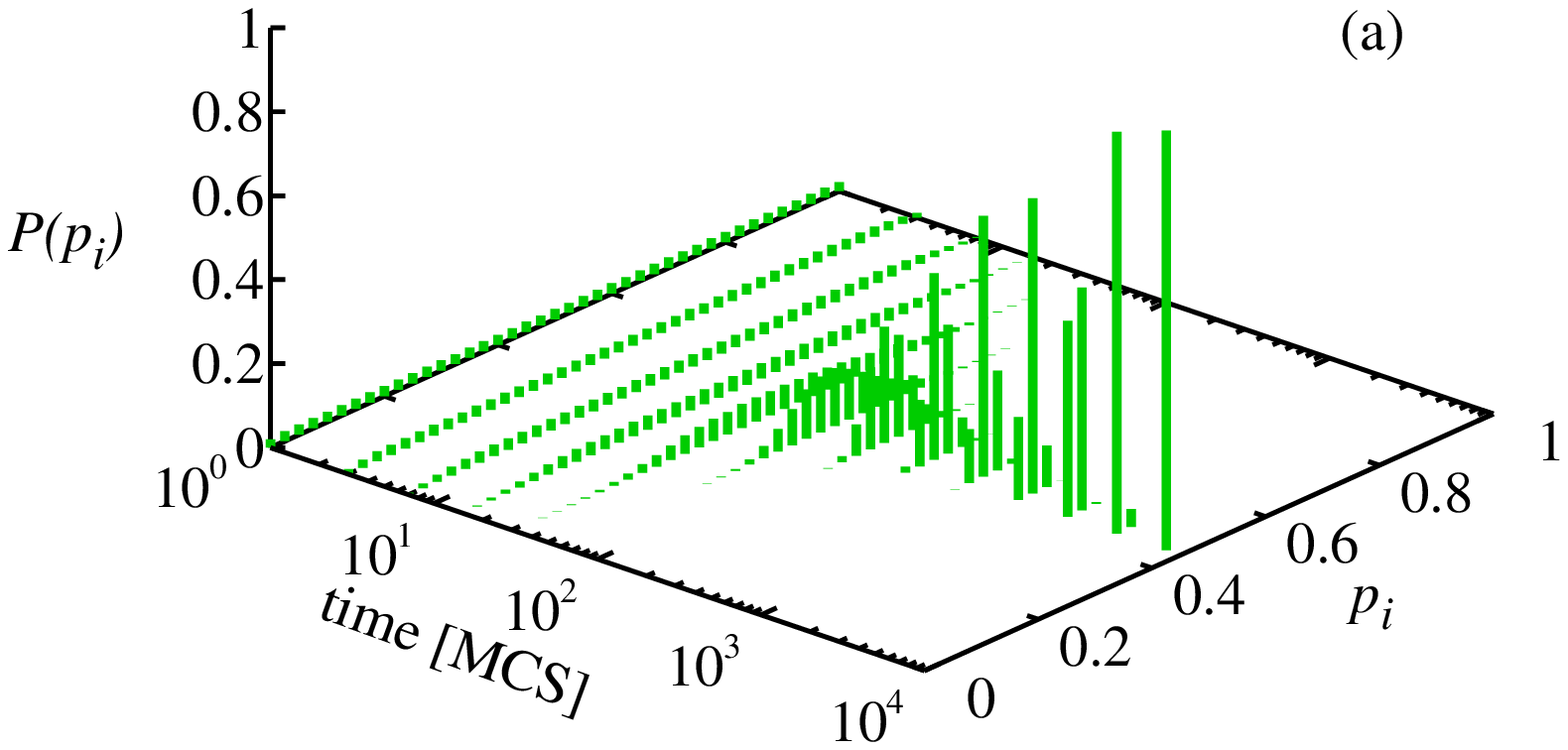}\\
\includegraphics[width=7.0cm]{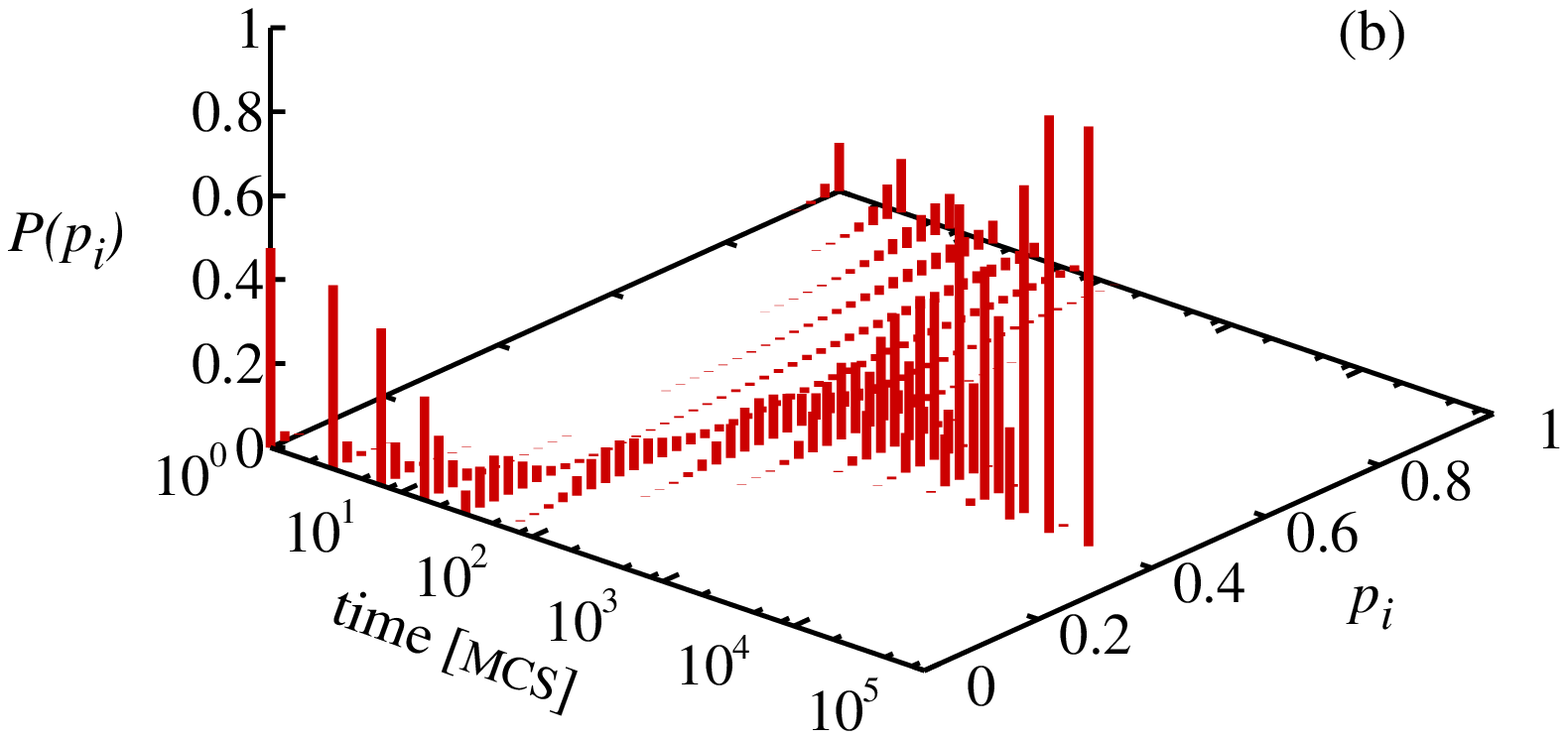}\\
\caption{Time evolution of strategy distributions started from a uniform state, shown in panel (a), and from a two-state initial state where only extreme strategies are present, shown in panel (b). In both cases gradual imitation rule is applied and the system terminates into a homogeneous state where $p_i \approx 0.4$. The linear size of square greed is $L=800$ and the same $T=1.2$ temptation value is used for both cases.}\label{dist}
\end{figure}

One may raise that the final destination may depend sensitively on the initial state. Of course this effect is relevant when the available initial strategies are limited and they span only a finite interval of the whole $[0,1]$ profile. For instance, if we apply the initial state used in panel~(a) in Fig.~\ref{over_under} then the system is unable to leave these strategies because of the character of the gradual imitation rule. But this effect becomes irrelevant if extreme strategies are present initially, no matter other strategies are absent first. This phenomenon is illustrated in Fig.~\ref{dist} where we plotted the time evolution of strategy distribution for two different cases. In the first case, plotted in panel~(a), all possible strategies are present uniformly in the initial state. Later strategies die out gradually, first those who are close to the extreme strategies. Finally, only a single strategy survives and the system terminates onto an absorbing state.

When the evolution is launched from the state contains only the extreme strategies then the system also terminates into an absorbing state where only a single strategy prevails. This process is shown in Fig.~\ref{dist}(b), where the number of available strategies gradually increases and after the evolution of strategy-distribution becomes similar to the above discussed case.

\begin{figure}[h!]
\centering
\includegraphics[width=7.0cm]{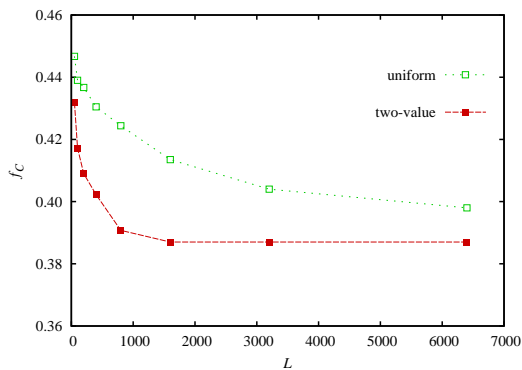}\\
\caption{Average cooperation level of final states in dependence of system size when the evolution is launched from a uniform and a two-strategy initial state on a square lattice. This plot suggests that the large size limit can be reached more easily when only extreme strategies are present in the initial state. Similarly to Fig.~\ref{dist} here $T=1.2$ and $N=20$ were applied. Results are averaged over 200 independent runs.}\label{fix}
\end{figure}

Naturally, the fixation interval, means which strategy survives in the end, may depend slightly on the system size. This is true especially for small system sizes and large $N$ values. This effect is illustrated in Fig.~\ref{fix} where we plotted the average of final cooperation levels in dependence of system size for both previously mentioned initial states.

Here we have averaged the cooperation levels of final states over several independent runs by using the same system size at fixed temptation value where the initial state was either uniform or two-strategy state of extreme strategies. This comparison suggests that albeit the fixation values tend to the same level in the large-scale limit, but the convergence in the two-strategy case is much faster. Practically, in the latter case we can obtain the values valid in the large size limit already at $L=1000$ linear size.

\begin{figure}
\centering
\includegraphics[width=7.0cm]{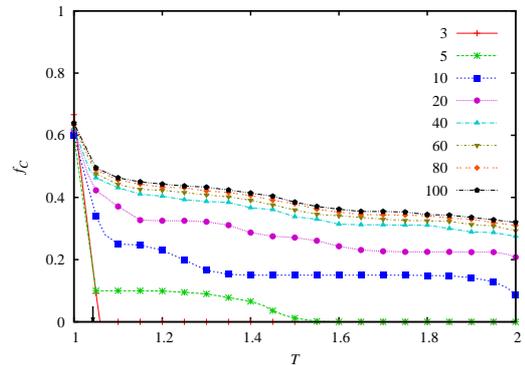}\\
\caption{Cooperation level in dependence on temptation parameter on square-lattice for different values of $N$ as indicated in the legend. Arrow indicates the highest temptation value where cooperators can survive in the traditional two-strategy model. Lines are just guide to the eye. Linear system sizes are $L=1000$.}\label{sqr}
\end{figure}

In the following we present our key observation about the impact of gradual learning protocol obtained on square lattice interaction graph. As Fig.~\ref{over_under} already signaled, the application of this strategy update rule can elevate the cooperation level in a population. This effect is confirmed by Fig.~\ref{sqr} where we summarized the cooperation level in dependence of temptation parameter for different $N$ values. These plots suggest that gradual learning is beneficial for cooperator strategies and the supporting mechanism becomes more visible for large $N$ values. Since large $N$ involves just a tiny step toward teacher's strategy, hence we can conclude that the more careful strategy change is applied, the stronger cooperation supporting effect is reached. To estimate properly this remarkable effect, in this plot we have also marked by an arrow the critical temptation value where cooperators die out in the traditional two-strategy model. 

\begin{figure}
\centering
\includegraphics[width=7.0cm]{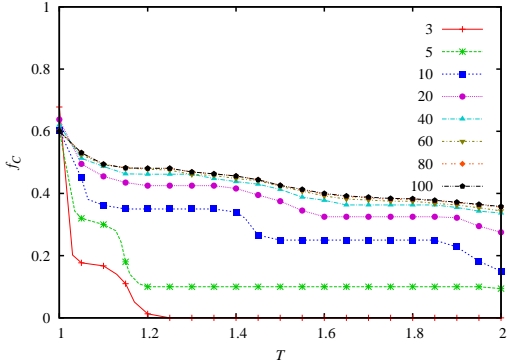}\\
\caption{Cooperation level in dependence on temptation parameter on random graph where we used the same $k=4$ degree number as for square lattice. The applied $N$ values are marked in the legend. Lines are just guide to the eye. Systems contain $6.4 \cdot 10^5$ players and results are averaged over 20 independent runs.}\label{rrg}
\end{figure}

We have also checked the robustness of this effect on other kind of interaction graphs. By leaving the translation invariant symmetry of lattice structures we first study a random graph which exhibits small-world character. To allow fair comparison with the results of square lattice we used the same $k=4$ degree distribution for each node \cite{szabo_jpa04}. Figure~\ref{rrg}, where our results are summarized, suggests that similarly high cooperation supporting effect can be observed, hence our observation is not restricted to lattice structures, but can also be detected for random graphs.  

\begin{figure}[h!]
\centering
\includegraphics[width=7.0cm]{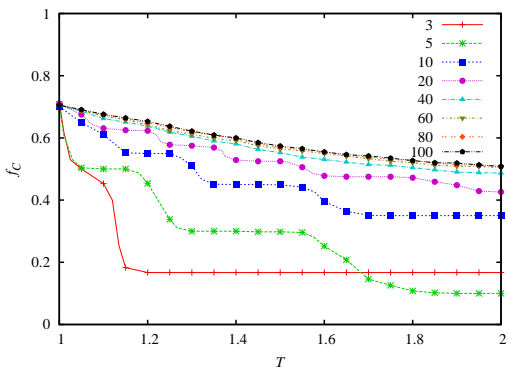}\\
\caption{Cooperation level in dependence on temptation parameter on scale-free networks where $\langle k \rangle =4$ is applied for different values of $N$ as indicated in the legend. Lines are just guide to the eye. Systems contain $10^5$ players and results are averaged over 50 independent runs.}\label{sf}
\end{figure}

Last, we present our results obtained for scale-free interaction graphs. This topology represents a separate class of graphs where the degree distribution of nodes is highly heterogeneous \cite{barabasi_s99}. To avoid additional effect, here we keep the $k=4$ average degree level and normalize the players' payoff by their degree value. In this way we can detect the pure impact of gradual learning protocol on cooperation level without disturbed by the predominant impact of accumulated payoff values on strongly heterogeneous graph \cite{santos_prl05}. The results, summarized in Fig.~\ref{sf}, highlight that heterogeneous degree distribution does not block the positive impact of gradual learning. On the contrary, the highest improvement can be detected here. For example, cooperators survive already for $N=3$ case in the whole $T$ interval, where learning is less gradual. Furthermore, players remain mostly in cooperator state even for high $T=2$ value in the large $N$ limit.

Previous works revealed the highly asymmetric way of strategy propagation for cooperator and defector strategies in spatial systems \cite{szabo_pr07, szolnoki_pa08, santos_jeb06, gomez-gardenes_prl07}. While defectors invade their neighborhood fast and easily, the alternative strategy needs to build up a "phalanx" of cooperators \cite{sigmund_10}, which is a slower process. Since the gradual learning strategy update rule may modify these dynamical processes, it is instructive to explore its direct consequences on the mentioned propagation courses. 

For this reason we used a prepared initial state where only extreme defector and cooperator strategies form separate domains and monitored the front propagation speed by measuring the domain growth of dominant strategy. We have repeated it at different $N$ values by using two characteristic temptation values. In the traditional model $T=0.9$ provides a clear dominance for cooperators while $T=1.1$ ensures that defectors prevail. Using these as references we can measure how mentioned domains grow for different $N$ values. As we already noted, the usage of different $N$ means different level of gradual learning protocols. In this way we can compare directly how the introduction of intermediate strategies modify the invasion process. 

\begin{figure}[h!]
\centering
\includegraphics[width=7.0cm]{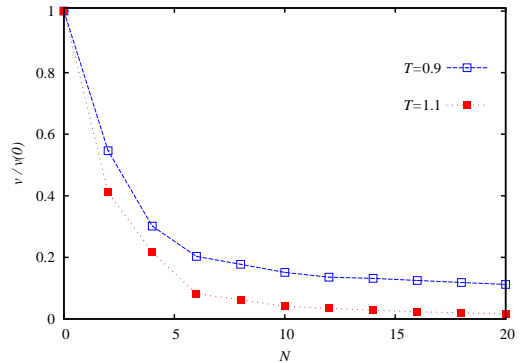}\\
\caption{Invasion speed of dominating strategies compared to the value of the classic two-strategy model in dependence of the number of inner competing strategies. During the imitation process players apply gradual learning. In case of $T=0.9$ cooperation, while for $T=1.1$ defection prevails when competition is released.}\label{speed}
\end{figure}

In Figure~\ref{speed} we compare how the speed of invasions changes for the two characteristic temptation values by increasing the number of intermediate states. Note that $N=0$ corresponds to the traditional two-strategy model therefore the plotted $v /v(0)$ represents the relative change of invasion speed. As expected, the introduction of gradual learning will decelerate the propagation processes for both cases. In other words, both the propagation of successful defector state and successful cooperator state will slow down. However, there is significant difference, because the modified learning protocol hinders the propagation of defection more efficiently. This biased impact on propagations explain why a strategy-neutral update rule has a cooperator supporting net consequence on the competition of strategies.

This argument is in nice agreement with our previous observations because the fastest invasion of defector state was previously reported in heterogeneous graphs. Since gradual learning protocol blocks this propagation efficiently therefore the largest increment in cooperation level can be expected for such interaction graphs.

\section{Conclusion}

Motivated by previous reports about microscopic dynamics sensitive outcomes of evolutionary process in present work we have explored the possible consequences of modified imitation rule protocols. Instead of accurate adoption we have introduced exaggerated and gradual imitation rules. While in the former protocol a learner enlarges the strategy change between the learner's and master's strategies, in the latter protocol the learner makes only a minimal step toward master's strategy during the update process. The former choice can be interpreted that the learner wants an even higher payoff than the one collected by the master player. The latter protocol mimics the situation when the learner is more careful and is reluctant to make drastic strategy change. These protocols can only be executed if we introduce a multi-strategy system where intermediate level of cooperation is also possible.

We have observed that the exaggerated update protocol has no particular role on the evolutionary process and eventually the system terminates onto the traditional two-strategy model where only always defect and always cooperate players are present. The gradual learning protocol, however, may influence the evolution significantly. If we make the strategy change tiny by introducing large number of intermediate strategies then the cooperation level can be elevated remarkably. We note that a conceptually similar observation was made in \cite{jimenez_epjb09} where the authors introduced a weighted average as the way to update investments.

We have demonstrated that the cooperation supporting mechanism of gradual learning can be explained by a biased impact on strategy propagation process. Indeed, gradual learning decelerates the invasion for both main strategies, but the blocking of defector invasion is more effective. In this way a strategy-neutral microscopic rule has a biased impact on the competition of strategies. We note that this observation fits nicely to the general concept when there is a non-trivial consequence of a neutral intervention on strategy evolution \cite{szolnoki_pre09, fu_mj_pa19, szolnoki_epl14b, liu_yh_c19, shao_yx_epl19, szolnoki_pre18, xu_zj_c19}.  

Our observations are robust and remain valid by using different types of interaction graphs. No matter translation invariant lattice, or random, or highly heterogeneous topology is used, the positive consequence of the gradual learning protocol is remarkable. We note, however, that similar cooperator supporting effect is expected on interdependent or multilayer networks \cite{wang_z_epl12, boccaletti_pr14, jiang_ll_srep15, allen_pre18, yang_gl_pa19}. This expectation is based on previous results which already highlighted the decisive role of propagation process in such complex networks \cite{szolnoki_njp13, arruda_pr18}.

\vspace{0.5cm}

This research was supported by the Hungarian National Research Fund (Grant K-120785) and by the National Natural Science Foundation of China (Grants No. 61976048 and No. 61503062).

\end{document}